\documentclass[12pt]{iopart}
\usepackage[utf8]{inputenc}
\usepackage[T1]{fontenc}
\pdfoutput=1 
\usepackage{amssymb}
\usepackage{iopams}
\expandafter\let\csname equation*\endcsname\relax
\expandafter\let\csname endequation*\endcsname\relax
\usepackage{amsmath}
\usepackage{color}
\usepackage{graphicx}
\usepackage{float}
\usepackage{hyperref}
\hypersetup{colorlinks=true, citecolor=blue, urlcolor=blue, linkcolor=blue}

\begin{document}

\title{Double moir\'{e} with a twist: super-moir\'e in encapsulated graphene}

\author{M. An\dj elkovi\'{c}, S. P. Milovanovi\'{c}, L. Covaci, and F. M. Peeters}

\address{Departement Fysica, Universiteit Antwerpen, Groenenborgerlaan 171, B-2020 Antwerpen, Belgium}
\ead{misa.andelkovic@uantwerpen.be, slavisa.milovanovic@uantwerpen.be, lucian.covaci@uantwerpen.be, francois.peeters@uantwerpen.be}
\vspace{10pt}
\begin{indented}
	\item[]\today
\end{indented}

\begin{abstract}
	A periodic spatial modulation, as created by a moir\'e pattern, has been extensively studied with the view to engineer and tune the properties of graphene. 
	Graphene encapsulated by hexagonal boron nitride (hBN) when slightly misaligned with the top and bottom hBN layers experiences two interfering moir\'e patterns, resulting in
	a so-called super-moir\'e (SM). This leads to a lattice and electronic spectrum reconstruction. A geometrical construction of the non-relaxed SM patterns allows us to indicate
	qualitatively the induced changes in the electronic properties and to locate the SM features in the density of states and in the conductivity. To emphasize the effect of lattice relaxation, we report band gaps at all Dirac-like points 
	in the hole doped part of the reconstructed spectrum, which are expected to be enhanced when including interaction effects. Our result is able to distinguish 
	effects due to lattice relaxation and due to the interfering SM and provides a clear picture on the origin of recently experimentally observed effects in such 
	trilayer heterostuctures.
\end{abstract}

\maketitle

\section{Introduction}
\label{Sec:Introduction}

	Encapsulation of graphene by hBN is a widely used technique for obtaining devices with 
	exceptional electronic properties~\cite{mayorov2011micrometer, kretinin2014electronic}. This is due to the fact that hBN serves as a protective layer for graphene, 
	isolating it from the contaminants from the environment, and to flatten the graphene layer. Although graphene and hBN have similar crystal structure 
	(difference between lattice constants is $\approx 1.8\%$), hBN is an inert band insulator that does not alter the electronic properties of graphene when 
	the two layers are misoriented. In contrast, when different lattices are closely aligned, spectrum~\cite{yankowitz2012emergence} and surface~\cite{woods2014commensurate} 
	reconstruction becomes significant as a consequence of the moir\'e periodicity between the two or more crystals. Moir\'e reconstruction 
	further	induces effects that are of great importance in the light of gap engineering and insulating phases~\cite{hunt2013massive, chen2014observation}, 
	tuning between metallic and superconductive phases~\cite{chen2019signatures} and even exploring interaction effects~\cite{chen2019evidence} in hBN/gr 
	multilayers. In addition, recent experimental advancement showed that it is possible to align two monolayer 
	sheets to a high rotation precision of $0.1^\circ-0.2^\circ$~\cite{kim2017tunable, cao2018unconventional}. Even dynamical tuning of the rotation angle 
	with precision below $0.2^\circ$ has been demonstrated~\cite{RibeiroPalau690dynamical_twisting}. 
	Further, high rotation controllability led to encapsulated graphene devices almost aligned to both hBN layers, gaining attention 
	when additional spectrum reconstruction was reported~\cite{Lujun_2019, finney_tunable_2019, Zihao2019}. 
	
	If a moir\'e pattern, formed by e.g. two monolayers, is brought in contact with another monolayer, the interference will lead to a super-moir\'e (SM) structure,
	leading to additional Dirac-like points. In this report we present a detailed study on the origin of the SM features, the resulting effects on the electronic 
	properties and on the lattice reconstruction. Even in the rigid form, small misorientation in the trilayer creates reconstructed bands that are different from 
	those found in bilayer graphene/hBN systems. We model the effect as a function of the relative twist between the layers and compare the results against a simple, 
	geometrical description of the moir\'e systems. Further, we discuss the renormalization of the Fermi velocity and the appearance of bands 
	with almost vanishing dispersion. Lastly, we report experimentally detectable band gaps at all Dirac-like points induced by the moir\'e patterns as a 
	result of lattice relaxation. Moreover, the used methodology can be applied to different almost aligned multilayer-heterostructures 
	with small difference in lattice constants of individual layers where similar effects can be expected.
	
	The paper is organized as follows. In section~\ref{Sec:Geometrical_consideration} geometrical properties of hBN/gr/hBN are described and 
	different SM spatial periodicities are derived. Section~\ref{Sec:Model_and_methodology} introduces the non-interacting tight-binding model for
	rotated gr/hBN heterostructures. Further, in section~\ref{sec:electronic_properties} the effect of the close alignment on the electronic 
	properties is examined, which is followed by discussions on the relaxation effects in section~\ref{Sec:relaxation}. Conclusions of the paper are summarized
	in section~\ref{sec:conclusion}.

\section{Geometrical considerations in rigid trilayers} 
\label{Sec:Geometrical_consideration}
	 
\begin{figure}[H]
		\begin{center}
			\includegraphics[width=\linewidth]{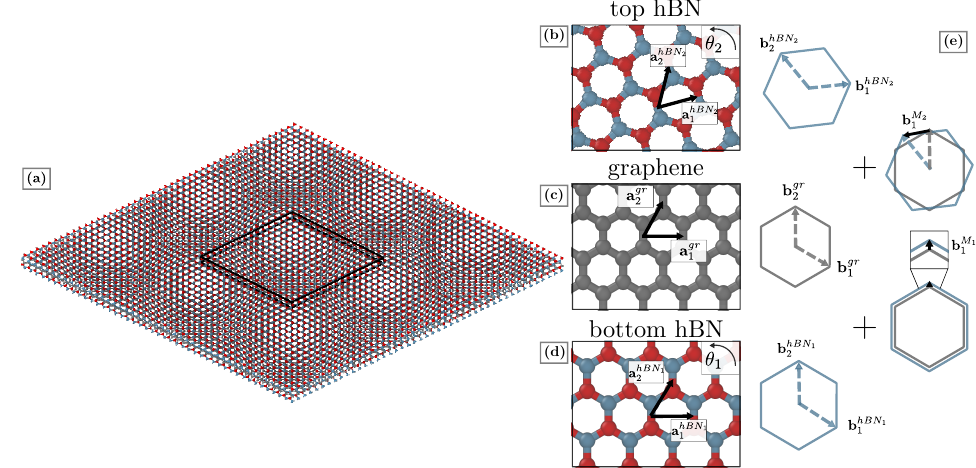}
			\caption{(a) Schematic of trilayer hBN/gr/hBN structure. Bottom and top hBN layers are twisted over $\theta_1=0^\circ$ and $\theta_2= 60^\circ$, respectively.  
			In panels (b - d) top hBN ($\theta_2$), graphene, and bottom hBN ($\theta_1=0^\circ$) lattices and corresponding unit cell vectors are shown. (e) Brillouin zone of each lattice, 
			and the moir\'e reciprocal vector between graphene and top/bottom hBN are depicted. The ratio between the two lattices is exaggerated 
			$\delta=16/15 - 1$ for visualization purposes.}
			\label{fig_1:Schematic}
		\end{center}
	\end{figure}

	Before further exploring the effects of the alignment between graphene and multiple hBN layers, a qualitative description of the moir\'e pattern
	is worth reviewing. In its rigid form, graphene and hBN can be defined as bipartite honeycomb lattices, as illustrated in Fig.~\ref{fig_1:Schematic}, with unit cell vectors of 
	graphene $\mathbf{a}^{gr}_{1}=a[1,0]$, $\mathbf{a}^{gr}_{2}=a[1/2, \sqrt{3}/2]$, and bottom ($i=1$) and top ($i=2$) hBN 
	$\mathbf{a}^{hBN_i}_{1,2}=\mathbf{R}(\theta_i)\mathbf{a}^{gr}_{1,2}/(1+\delta_{i})$, where $a\approx 0.246$ nm is the length of graphene unit cell, 
	$\mathbf{R}(\theta_i)$ represents a rotation matrix in counter-clock wise direction by an angle $\theta_i$, and $\delta_{1,2}$ are the relative ratios 
	between graphene and bottom/top hBN lattice constants, with the unrelaxed value of $56/55-1\approx 0.018$. Reciprocal vectors are defined as 
	$\mathbf{b}^{gr}_{1}=2\pi/a[1, -\sqrt{3}/3]$, $\mathbf{b}^{gr}_{2}=2\pi/a[0, 2\sqrt{3}/3]$, 
	$\mathbf{b}^{hBN_i}_{1,2}=\mathbf{R}(\theta_i)\mathbf{b}^{gr}_{1,2}/(1+\delta_{i})$ satisfying 
	$\mathbf{b}^{gr}_j \mathbf{a}^{gr}_k=\mathbf{b}^{hBN_i}_j \mathbf{a}^{hBN_i}_k=2\pi\delta_{j,k}$.

	From there, the top and the bottom moir\'e reciprocal vectors can be defined as a difference between the reciprocal vectors of graphene and hBN, 
	$\mathbf{b}^{M_i}_{1,2}=\mathbf{b}^{gr}_{1,2} - \mathbf{b}^{hBN_i}_{1,2}$. 
	
	The massless Dirac-Weyl equation describes the behaviour of a single graphene layer, while hBN layers can be modeled as 
	an effective potential and a gauge field in the continuum approximation~\cite{Moon_2014}. Still, in the continuum description strain and 
	relaxation effects are fitted parametrically, which results in loosing the generality of the model. This is why in the following a real-space representation of 
	a tight-binding Hamiltonian will be used. In a similar manner as in the continuum description, hBN substrate can be effectively modeled as a modification of 
	both hopping terms and the onsite energy. Following the work in~\cite{Diez_2014}, but considering only an additional potential that respects the periodicity 
	of the underlying moir\'e patterns, the change of the onsite energy is given by 
	\begin{equation} 
		\begin{split} 
			\varepsilon_{i,+}^{m, n} &= c_1(\theta_i,\delta_i) \left[ \cos(\mathbf{g}^i_1 \mathbf{r}_{m,n}) + 
			\cos(\mathbf{g}^i_3 \mathbf{r}_{m,n}) + \cos(\mathbf{g}^i_5 \mathbf{r}_{m,n}) \right],   \\
			\varepsilon_{i,-}^{m, n} &= c_2(\theta_i,\delta_i) \left[ \sin(\mathbf{g}^i_1 \mathbf{r}_{m,n}) + 
			\sin(\mathbf{g}^i_3 \mathbf{r}_{m,n}) + \sin(\mathbf{g}^i_5 \mathbf{r}_{m,n}) \right], 
		\end{split} 
		\label{eq:sublattice_potential}  
	\end{equation}
	where the parameters $c_1$ and $c_2$ contain the lattice information of the two crystals and the rotation angle between them, 
	$m$ and $n$ are indexes of the unit cell, $\mathbf{r}_{m,n}=m \mathbf{a}^{gr}_{1} + n \mathbf{a}^{gr}_{2}$, $\mathbf{a}_{1,2}$. The two 
	terms in Eq.~(\ref{eq:sublattice_potential}), $\varepsilon_{i,+}$ and $\varepsilon_{i,-}$ denote sublattice symmetric and antisymmetric potential, 
	respectively (due to hBN layer $i$). Rotated moir\'{e} reciprocal lattice vectors 
	$\mathbf{g}^i_j$ are $\mathbf{g}^i_j = \mathbf{R}\left((j-1)\pi/3\right)\mathbf{b}^{M_i}_1$.

\begin{figure}[H]
		\begin{center}
			\includegraphics[width=\linewidth]{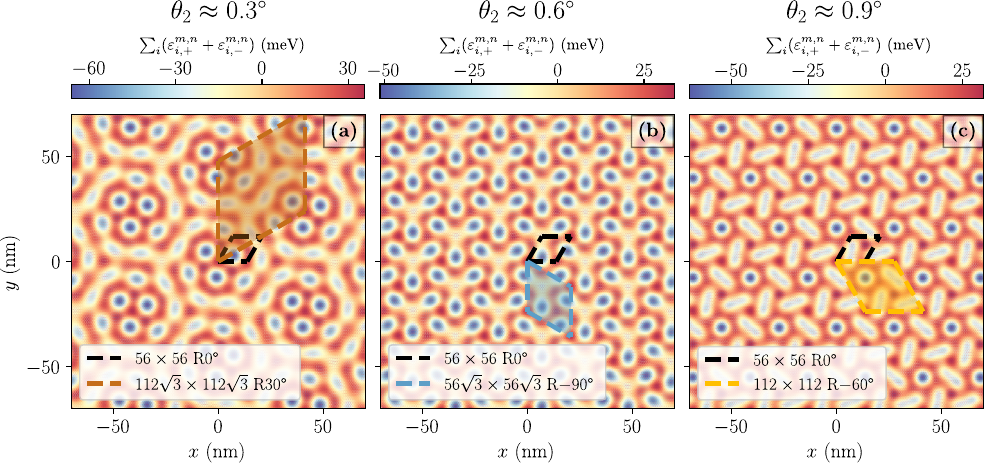}
			\includegraphics[width=\linewidth]{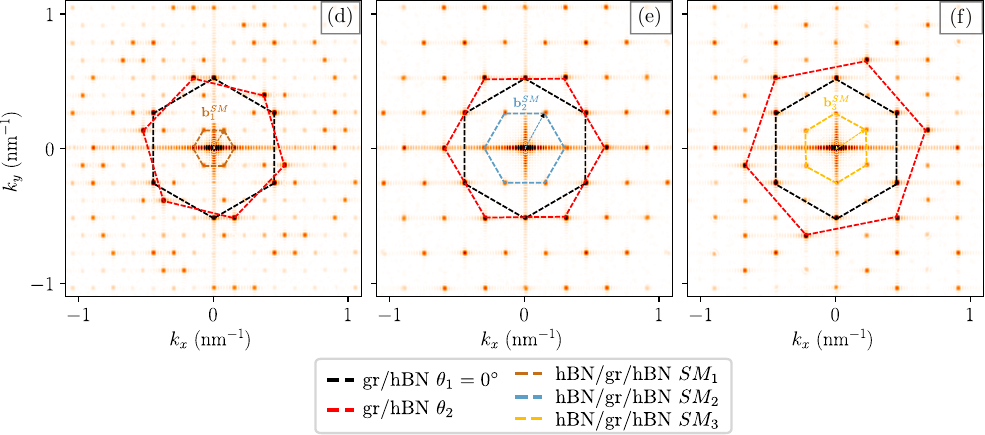}
			\caption{(a - c) Real space distribution of the potential in graphene defined in Eq.~(\ref{eq:sublattice_potential}) due to two hBN layers for 
			$\theta_1=0^\circ$ with top hBN layer twisted over $\theta_2\approx 0.3^\circ$, $\theta_2\approx 0.6^\circ$ and $\theta_2\approx 0.9^\circ$. 
			The parallelograms are the aligned moir\'e (black) and the different SM unit cells, defined in Wood's notation.
			(d - f) FT of the potential shown in panels (a - c), from which super-moir\'e period can be obtained. 
			Hexagons show the mini-Brillouin zones of aligned (black), twisted (red) and different super-moir\'es.}
			\label{fig_2}
		\end{center}
	\end{figure}

	In the case of almost aligned graphene with multiple hBN layers, the effective onsite terms will become superposition 
	of the changes due to each of the hBN layers, taking into account the exponential decrease of the interlayer interaction. Without loss of generality, it is 
	further assumed (except when considering the effect of the misalignment) that graphene is perfectly aligned with the bottom hBN layer, while the top 
	hBN is twisted. Taking this into account, in Fig.~\ref{fig_2} the superimposed onsite potential in graphene layer, 
	and the Fourier transformation of this potential, originating from the encapsulation, are shown. 
	From both the real and the reciprocal space it is clear that in addition to the moir\'e pattern due to aligned bottom hBN (marked in Fig.~\ref{fig_2} with 
	black unit cell in real space, and black mini-Brillouin zone in reciprocal space with $\lambda\approx 13.9$ nm), and the rotated moir\'e 
	between graphene and top hBN (red colour), there exist additional periodicities of longer range. The Fourier transformation (FT) in reciprocal 
	space further reveals which components are important. From this qualitative analysis, a comprehensive description of possible moir\'e patterns 
	are obtained. Straightforward geometrical consideration shows that beside the aligned and the rotated graphene-moir\'e spots in the FT, different 
	combinations of the two result in additional SM patterns, similar to the case of two-layered systems with hexagonal 
	symmetry~\cite{Zeller_2014}. The corresponding FT harmonics can be derived as translation (difference) vectors that match the two of the original 
	moir\'e patterns. Hence, the SM reciprocal vectors can be obtained as a difference between the aligned and rotated reciprocal vectors, 
	or their multiples.  
	
\begin{figure}[H]
		\begin{center}
			\includegraphics[width=\linewidth]{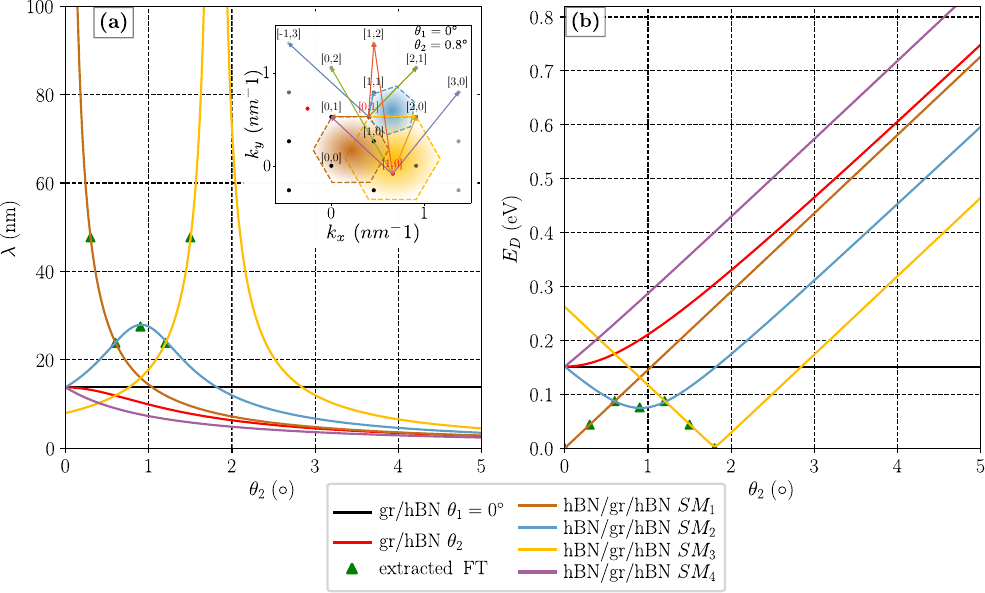}
			\caption{Change of the (a) moir\'e and super-moir\'{e} period and (b) the energy of the additional Dirac-like points as a function of rotation angle 
			$\theta_2$. Different lines represent the super-moir\'e harmonics defined in the inset of panel (a), the aligned graphene-hBN 
			($\theta_1=0^\circ$) and the rotated graphene-hBN moir\'e ($\theta_2$). Super-moir\'{e} unit cell lengths extracted 
			from FT in Fig.~\ref{fig_2} and two additional angles ($\theta_2\approx1.2^\circ~\text{and}~1.5^\circ$) are shown with 
			green triangles.}
			\label{fig_3}
		\end{center}
	\end{figure}

	The real space description of the SM starts with previously introduced graphene and hBN lattices and the corresponding moir\'e patterns. From there, an arbitrary 
	super-moir\'e vector is
	\begin{equation}
		\mathbf{b}^{SM} =\sum \left( \begin{array}{cc}
			i  &  j \\
		   -k  & -l 
		 \end{array} \right) \odot
		 \left( \begin{array}{cc}
		   \mathbf{b}^{M_1}_{1} & \mathbf{b}^{M_1}_{2} \\
		   \mathbf{b}^{M_2}_{1} & \mathbf{b}^{M_2}_{2}
		 \end{array} \right),\\
		 \label{eq:define_arb_super_moire}
	\end{equation}
	where $\odot$ denotes the element-wise product with $i, j, k, l \in \mathbb{Z}$. 
	This frequency analysis leads to analytical expressions for possible SM patterns obtained for an arbitrary rotation between the two 
	moir\'{e} patterns. The shortest reciprocal (or the longest real space) SM vectors are the ones affecting the low energy spectrum. 
	Different SM harmonics at low rotation angles which emerge as the difference between the moir\'e harmonics of different order agree well 
	with the experimentally observed features from~\cite{Zihao2019}. These are shown in the inset of Fig.~\ref{fig_3}, where
	the coloured hexagons represent their mini-Brillouin zones. Under the assumptions of $\theta_1=0^\circ$, and $\delta_1=\delta_2=\delta$, four different 
	low-energy, low-angle harmonics, which will be later used for comparison against the numerical results, and the corresponding real space 
	periodicities $\lambda^{SM}_i$ (using relation $\lambda=4\pi/\sqrt{3}/|\mathbf{b}_i|$) are
	\begin{equation}
		\begin{alignedat}{3}
			\lambda_1^{SM}&=\frac{a(1+\delta)}{\sqrt{2-2\cos(\theta_2)}}, &&\quad \mathbf{b}^{SM}_1=\sum \left( \begin{array}{cc}
				 0  &  1 \\
				-0  & -1 
			  \end{array} \right) &&\odot
			  \left( \begin{array}{cc} 
				\mathbf{b}^{M_1}_{1} & \mathbf{b}^{M_1}_{2} \\
				\mathbf{b}^{M_2}_{1} & \mathbf{b}^{M_2}_{2}
			  \end{array} \right),\\
			\lambda_2^{SM}&=\frac{a(1+\delta)}{\sqrt{(2-\delta)(1-\cos(\theta_2))+\delta^2-\sqrt{3}\delta\sin(\theta_2)}}, &&\quad \mathbf{b}^{SM}_2=\sum \left( \begin{array}{cc}
				 0  &  1 \\
				-1  & -1 
			\end{array} \right) &&\odot
			\left( \begin{array}{cc} 
			  \mathbf{b}^{M_1}_{1} & \mathbf{b}^{M_1}_{2} \\
			  \mathbf{b}^{M_2}_{1} & \mathbf{b}^{M_2}_{2}
			\end{array}\right),\\
			\lambda_3^{SM}&=\frac{a(1+\delta)}{\sqrt{2+3\delta^2-2\cos(\theta_2)-2\sqrt{3}\delta\sin(\theta_2)}}, &&\quad \mathbf{b}^{SM}_3=\sum \left( \begin{array}{cc}
				0   & 1 \\
				-2  & 0
			\end{array} \right) &&\odot
			\left( \begin{array}{cc} 
			  \mathbf{b}^{M_1}_{1} & \mathbf{b}^{M_1}_{2} \\
			  \mathbf{b}^{M_2}_{1} & \mathbf{b}^{M_2}_{2}
			\end{array} \right),\\
			\lambda_4^{SM}&=\frac{a(1+\delta)}{\sqrt{(2+\delta)(1-\cos(\theta_2))+\delta^2+\sqrt{3}\delta\sin(\theta_2)}}, &&\quad \mathbf{b}^{SM}_4=\sum \left( \begin{array}{cc}
				 1  &  0 \\
				-0  & -1 
			\end{array} \right) &&\odot
			\left( \begin{array}{cc} 
			  \mathbf{b}^{M_1}_{1} & \mathbf{b}^{M_1}_{2} \\
			  \mathbf{b}^{M_2}_{1} & \mathbf{b}^{M_2}_{2}
			\end{array} \right),\\
		\end{alignedat}
		\label{eq:sm_harmonics}
	\end{equation} 
	as being shown in Fig.~\ref{fig_3}(a). Further, energy of the reconstruction is related to the period as $E_D = \pm 2\pi \hbar v_F/(\sqrt{3} \lambda)$ 
	and is shown in Fig.~\ref{fig_3}(b).
	Previous consideration also applies for any combination of 2D materials with honeycomb lattice. Notice that it is the close matching of the 
	lattice constants of the two materials that results in the reconstruction of both the energy dispersion through the formation of moir\'e induced 
	mini-bands and secondary Dirac points (SDPs) and the reconstruction of the surface due to the van der Waals interaction as further discussed in 
	section~\ref{Sec:relaxation}. 
	
	The qualitative analysis of multiple superimposed moir\'e patterns does not include 
	a quantitative description of the coupling between the layers. Experimental reports have shown that the van der Waals coupling leads to straining of the lattice,
	resulting in a commensurate state of the hBN/gr bilayers for almost aligned structures with angle smaller than $1^\circ$~\cite{woods2014commensurate}. 
	For larger rotation angles, the hBN potential only weakly affects the graphene layer due to the fast moir\'e oscillations~\cite{woods2014commensurate}. 
	It is only in the regime of highly coupled layers that the fine effect of mini-band formation will occur. This is why in the 
	following section, hBN/gr/hBN trilayers are simulated following the real space methodology, where the hBN layers are considered fully and not as an effective 
	potential, which captures the subtle SM and, after relaxing the sample, the strain effects. 
		
\section{Model and methodology}
\label{Sec:Model_and_methodology}
	 
	We start from the graphene-hBN non-interacting TB Hamiltonian~\cite{Moon_2014}
	\begin{equation}
		H = -\sum_{i,j}t(\boldsymbol{r}_i,\boldsymbol{r}_j)c_i^{\dagger}c_j + V_{i}(\boldsymbol{r}_i),
		\label{eq:Hamiltonian}
	\end{equation}
	with the hopping term between sites $i$ and $j$ defined based on the Slater-Koster like terms~\cite{Slater1954}
	\begin{equation}
		\begin{split}
			-t(\boldsymbol{r}_i,\boldsymbol{r}_j) =V_{pp\pi}\left[1-\left(\frac{\boldsymbol{d\cdot e}_z}{d}\right)^2
			\right] + V_{pp\sigma}\left(\frac{\boldsymbol{d\cdot e}_z}{d}\right)^2,
		\end{split}
	\end{equation}
	\begin{equation}
		\begin{split}
			V_{pp\pi} &= V_{pp\pi}^0e^{-\frac{d-a_{cc}}{r_0}}, \\
			V_{pp\sigma} &= V_{pp\sigma}^0e^{-\frac{d-c_0}{r_0}},
			\label{equation:hoppings}
		\end{split}
	\end{equation}
	with $V_{pp\pi}^0=-2.7eV$ and $V_{pp\sigma}^0=0.48eV$ are intralayer and interlayer hopping integrals, respectively. Carbon-carbon distance in
	graphene is $a_{cc}\approx 0.142$ nm, $c_0\approx 0.335$ nm is the interlayer distance between graphene and hBN layers, $\boldsymbol{d}$ 
	is the vector between two sites, $d$ is the distance between them, and $r_0 = 0.3187 a_{cc}$ is chosen in order to fit the next-nearest 
	intralayer hopping to $0.1V_{pp\pi}^0$. Onsite terms of boron and nitrogen species are set to $V_B=3.34~eV$  and $V_N=-1.40~eV$. 
	We have taken into account the next-next-nearest neighbour coupling in graphene, and the coupling within the radius of $1.5c_0$ in between graphene and 
	hBN layers.

	For obtaining electronic properties from the Hamiltonian defined in Eq.~(\ref{eq:Hamiltonian}) the kernel polynomial method (KPM)~\cite{Weie2006kernel} 
	implemented in Pybinding~\cite{Moldovan2017pybinding} is used. Band structure is calculated as the density of states (DOS) of a $\mathbf{k}$-dependent Hamiltonian, 
	where $\mathbf{k}$ takes its values along significant points in the super-moir\'e mini-Brillouin zone. The total DOS in the periodic system can be defined as $\mathrm{DOS}=1/N_k\sum_{k_i} \mathrm{DOS}(H(\mathbf{k}_i))$,
	where $\mathbf{k}_i$ takes values from the full super-moir\'e mini-Brillouin zone, and $N_k$ is the total number of points. 

	In general, the simulated structure for arbitrary rotation angles $\theta_1$ and $\theta_2$ is incommensurate, which means that the unit-cell is not properly defined. Commensurate lattice appears when an additional condition given by the Diophantine equation is fulfilled
	\begin{equation}
		\mathbf{L_{SM}} = i \mathbf{a}^{gr}_{1} + j \mathbf{a}^{gr}_{2} = k \mathbf{a}^{hBN_1}_{1} + l \mathbf{a}^{hBN_1}_{2} = m \mathbf{a}^{hBN_2}_{1} + n \mathbf{a}^{hBN_2}_{2},~\quad {i, j, k, l, m, n} \in \mathbb{Z},
		\label{eq:diophantine}
	\end{equation}
	together with $\left|\mathbf{L_{SM}}\right|=\lambda^{SM}$ as given in Eq.~(\ref{eq:sm_harmonics}). These conditions are only met for a discrete set of rotation angles, which are sparsely distributed at low twists. 
	To be able to obtain bulk properties of systems with an arbitrary rotation,
	angle dependent local density of states (LDOS), spatial distribution of the LDOS and the Kubo DC conductivity in the linear response regime, as presented 
	in~\cite{Garca2015}, are calculated using circular hBN/gr/hBN flakes with radius $R=150$ nm, where the effect of the boundaries is minimized by using absorbing boundary 
	conditions~\cite{Grozdanov_1995,Andelkovic2018,Munoz2015}. High resolution conductivity results are obtained using the single-shot algorithm~\cite{Ferreira2015}. For calculating the DC conductivity, 
	Anderson disorder (zero average onsite energy) with width of $W=0.1 V_{pp\pi}^0$ is considered. Typically 15000 KPM moments are computed, and Lorentz kernel is used for reconstructing desired quantities.

	We also performed molecular dynamics simulations (MD) for the double aligned structure for relaxing the structure. This was done through total energy minimization of 
	the unit cell, with certain constraints. We used the Brenner potentials for the graphene layer, Tersoff potentials for the B-N interaction in the top hBN and the 
	Morse potential developed in~\cite{Argentero2017} for the inter-layer interactions. Simulations were performed within the "Large-scale atomic.molecular massively 
	parallel simulator" (LAMMPS) software~\cite{lammps1, lammps2}. In order to mimic a thicker hBN substrate (in most of the experimental setups hBN substrate is usually a thick slab~\cite{dean2010boron})	
	we fixed the bottom hBN configuration, while only allowing the relaxation of the other two layers. The total energy was minimized until the error in forces calculated between 
	two iteration steps was below $10^{-6}$ eV/\AA.

\section{Evidence of the super-moir\'e in the electronic properties}
\label{sec:electronic_properties}
	
\begin{figure}[!htb]
		\begin{center}
			\includegraphics[width=\linewidth]{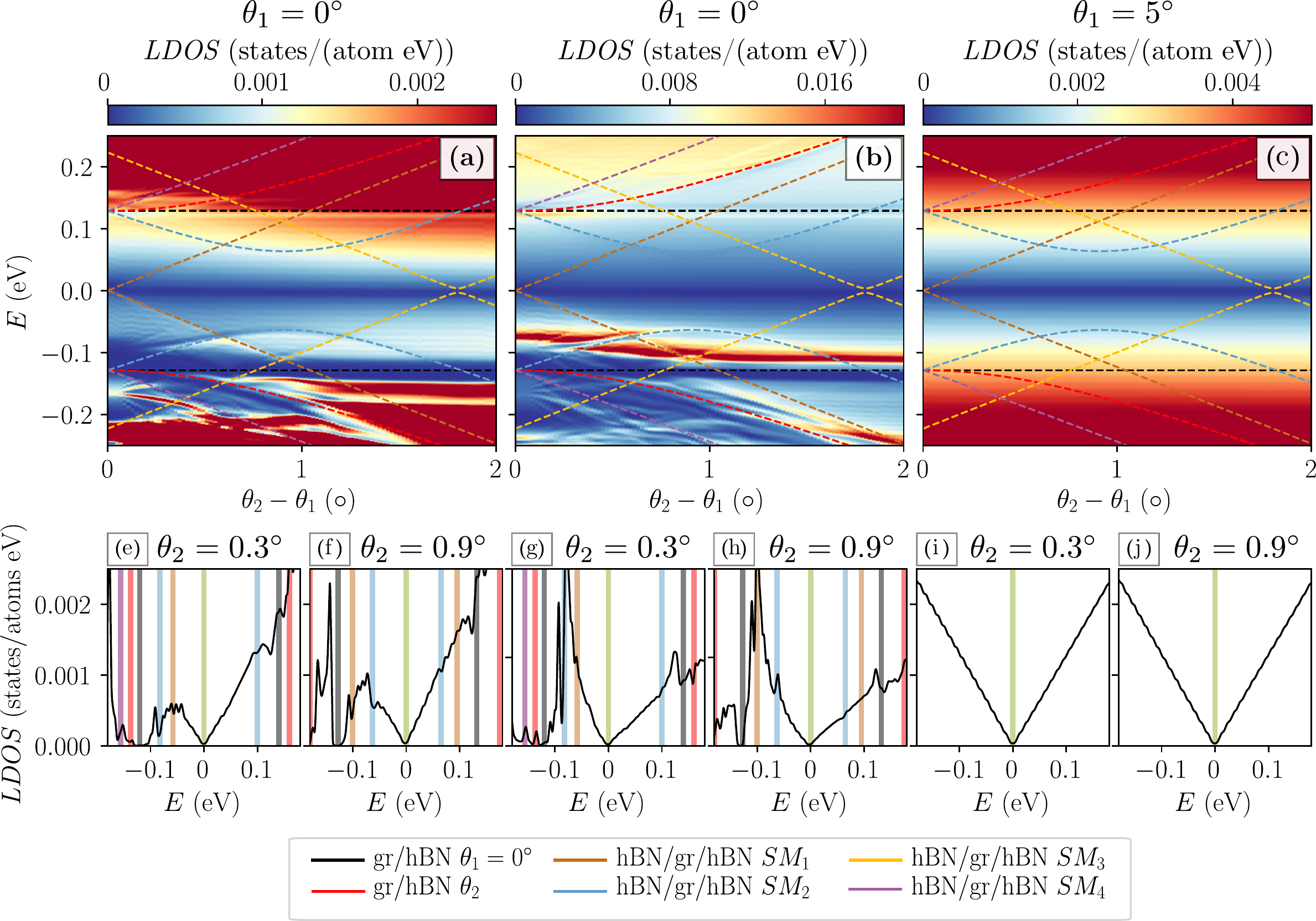}
			\caption{LDOS versus the angle of rotation and the energy. 
			(a) LDOS at a graphene site that sits on top of boron and (b) on top of a nitrogen atom from the bottom layer for rotation $\theta_2$ of the top hBN layer, keeping the bottom 
			layer aligned. (c) Shows the low energy LDOS at the same site as in (a), for the same relative rotation between the hBN layers ($\theta_2-\theta_1$), 
			while the graphene layer is misaligned for $\theta_1=5^\circ$. Other than moir\'e SDPs, additional harmonics from Eq.~(\ref{eq:sm_harmonics}) are shown with 
			dashed lines, using $v_F\approx0.75\times10^6$ m/s. Panels (e-j) show LDOS cuts at angles $\theta_2\approx 0.3^\circ$ and $\theta_2\approx 0.9^\circ$
			from panels (a-c), respectively. The fit of the dominant harmonic effects on the LDOS is shown with coloured stripes.}
			\label{fig_4}
		\end{center}
	\end{figure}

	To further investigate the behaviour of the hBN/gr/hBN system we consider all three layers, as well as coupling between them. All three layers are kept rigid, and 
	the lattice relaxation is discussed later in detail. The change of the LDOS with tuning the rotation angle is shown in Fig.~\ref{fig_4}. 
	In panels (a) the LDOS of a lattice site of graphene that sits on top of boron and in (b) for a graphene lattice site on top of a nitrogen atom from the bottom layer 
	are shown for low rotation angles. 
	In agreement with previous theoretical descriptions of bilayer gr/hBN samples~\cite{Moon_2014, Wallbank_2013}, strong asymmetry between electron and hole states 
	exists, with much stronger features on the hole doped side. This asymmetry can be traced back to the correlation between spatial variations of the onsite energy 
	and the variations in the effective hopping with neighboring atoms in the graphene layer~\cite{DaSilva2015el_hole_asymmetry}.

	Focusing on panels (a) and (b) of Fig.~\ref{fig_4}, one notices that different SM reconstruction points can be observed. 
	The band gap that occurs in LDOS between energies $\approx(-125, -135)$ meV (marked with black dashed line) remains constant in energy with rotation angle. 
	Converting the energy to the corresponding period gives $\lambda \approx 13.9$ nm, which can be addressed to the SDP from the aligned 
	graphene/hBN moir\'e pattern. The dip in the LDOS observed at energy lower than this one corresponds to the secondary Dirac point due to graphene 
	and rotated graphene hBN, and is marked with red dashed curve. Additionally, in the low energy regime, LDOS exhibits a series of peaks marked 
	with brown, blue and yellow dashed curves, again stronger on the hole doped side. Having in mind that these dips appear at lower energies than the
	SDPs due to aligned and rotated hBNs, they can be attributed to the SM periods. In fact, these are none other than angle-energy features 
	of three harmonics as defined by Eq.~(\ref{eq:sm_harmonics}) and shown in Fig.~\ref{fig_3}(b), which are now reproduced in all panels. 
	In contrast with the plot in Fig.~\ref{fig_2}(b), suppressed Fermi velocity of $v_F\approx0.75\times10^6$ m/s was used in order to obtain good agreement between
	the SM features in the LDOS and the derived harmonics, suggesting that the SM effect alters the electron velocity.
	
	The pronounced sublattice asymmetry when comparing the LDOS of the two graphene sublattices in panels (a) and (b) of Fig.~\ref{fig_4}  
	is resulting from the local stacking arrangement between graphene and the encapsulating hBN layers. As mentioned previously, important aspect 
	not considered in section~\ref{Sec:Geometrical_consideration} is the strength of the interlayer coupling. Weaker features on the 
	electron doped side and the disappearance of the SM features cannot be explained with former description. It is the effect of the interlayer 
	coupling that is reflected in the appearance of SM dips. Once the effective interlayer coupling is weakened through the effect of the 
	misalignment, similar as in the case of the	commensurate-incommensurate transition~\cite{woods2014commensurate}, the effect of the SM is lost. 
	To further explore the importance of close-alignment, Fig.~\ref{fig_4}(c) shows the LDOS of misaligned graphene layer of $\theta_1=5^\circ$ with the same 
	relative angle between the two hBN layers as in Figs.~\ref{fig_4}(a) and (b), and we see no signatures of SM. This explains why the SM effect is only recently 
	reported~\cite{Lujun_2019, finney_tunable_2019, Zihao2019} when one was able to achieve accurate control over the rotation angle between the layers, although hBN encapsulation has been a technique of choice for graphene transport 
	devices for a much longer period~\cite{mayorov2011micrometer}.

\begin{figure}[!htb]
		\begin{center}
			\includegraphics[width=\linewidth]{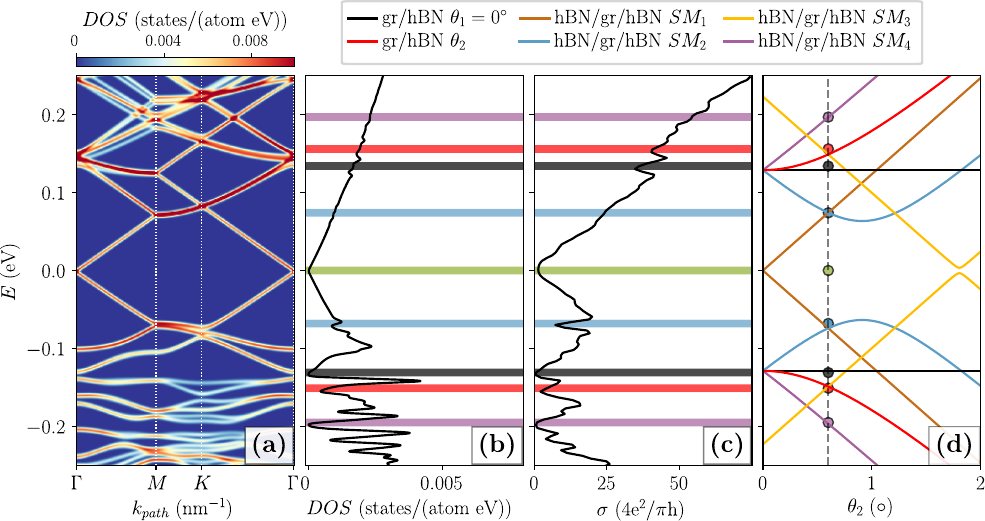}
			\caption{Electronic properties of hBN/gr/hBN system with $\theta_1= 0^\circ$ and $\theta_2\approx0.6^\circ$. (a) Band structure 
			along the $\Gamma - M - K - \Gamma$ path of the super-moiré mini-Brillouin zone. (b) DOS, and (c) conductivity. 
			Dominant super-moir\'{e} feature is marked with blue colour, black marks the aligned $\theta_1$ hBN/gr, and red 
			rotated $0.6^\circ$ hBN/gr features. Purple marks the higher order super-moir\'{e} feature. Green line marks the primary 
			graphene Dirac point. (d) Expected positions of the super-moir\'{e} features using $v_F\approx0.75\times10^6$ m/s. Different lines correspond to 
			harmonics defined in Eq.~\ref{eq:sm_harmonics}.}
			\label{fig_5}
		\end{center}
	\end{figure}

	In Fig.~\ref{fig_5} band structure, density of states (DOS), and the conductivity of the hBN/gr/hBN system with $\theta_2\approx 0.6^\circ$  
	are shown and compared against the expected energy of different moir\'es from section~\ref{Sec:Geometrical_consideration}. 
	Significant energies are marked with coloured stripes, each one corresponding either to the moir\'e or SM effect labeled in Fig.~\ref{fig_2}(c) and Fig.~\ref{fig_5}(d), 
	or to the primary Dirac point of graphene. Band structure and the corresponding DOS are important to unravel the different SM effects on the two types of carriers. 
	At positive Fermi energies (electron doped side), around $E\approx 70$ meV, a Dirac-like point appearing at the $\mathbf{K}$ point of the superlattice mini-Brillouin 
	zone is observed within the probing resolution. In contrast, the hole doped SM effect results in the appearance of band anti-crossing around $E\approx -70$ meV, 
	implying a different effect of the two features on the electronic properties. Further, due to the band folding, the moir\'e $\theta_{1,2}$ SDPs are appearing at the 
	$\Gamma$ and the $K$ point of the SM mini-Brillouin zone at energies $E_{SDP1}\approx \pm130$ meV and $E_{SDP2}\approx \pm150$ meV, 
	respectively. In between the two SDPs on the hole doped side, a band with reduced Fermi velocity appears, reflecting in a peak in the DOS. Due to its small 
	dispersion low conductivity can be observed. Compared against the SM effect in the DOS, transport properties reveal less pronounced electron-hole asymmetry, with 
	drops in the conductivity around the SM energy at both sides. In Fig.~\ref{fig_5}(d), different features are compared against the expected 
	energy at which the moir\'e harmonics affect the spectrum. Vertical line intersect different harmonics at angle $\theta_2\approx 0.6^\circ$, while the circular markers 
	are obtained from Fig.~\ref{fig_5}(b) and (c), once again confirming their origin. 

	In addition, further comparison between local properties - LDOS, and global - DOS and conductivity is given in the Supplementary information, section~\ref{sec:higher_rotation_angles}, 
	showing that the SM effect in transport vanishes already at rather small angles of $\theta_2\approx0.9^\circ$, different from the effect on the 
	LDOS shown in Fig.~\ref{fig_4}, where it was observed even for rotations larger than $1.5^\circ$. 
	 
\begin{figure}[!htb]
		\begin{center}
			\includegraphics[width=\linewidth]{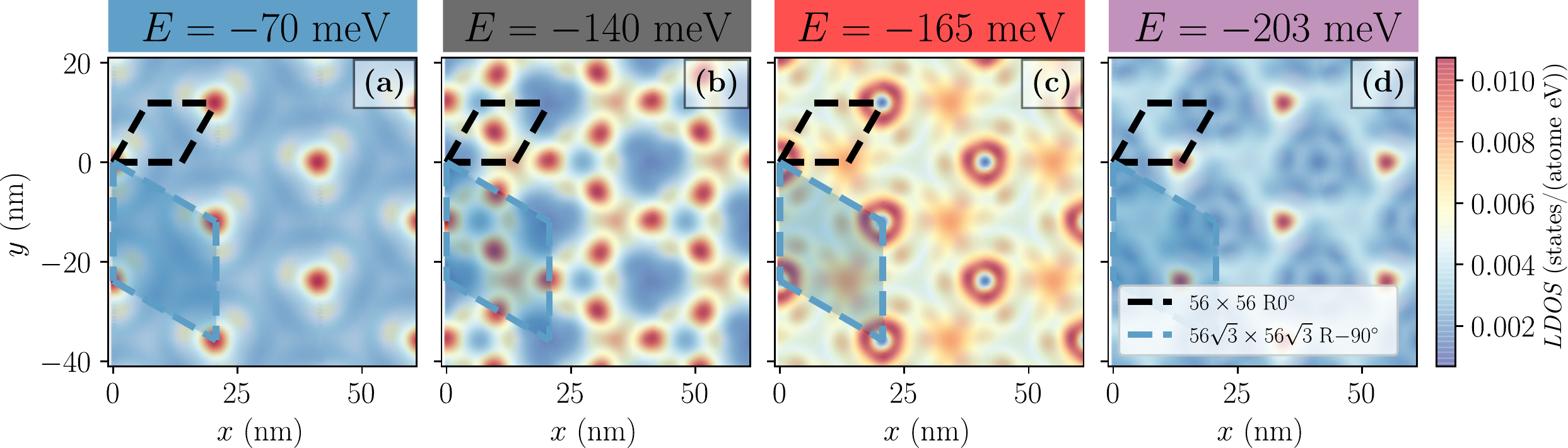}
			\caption{SLDOS maps of trilayer hBN/gr/hBN with $\theta_1= 0^\circ$ and $\theta_2\approx0.6^\circ$ at different energies near the features 
			shown with the same colours in Fig.~\ref{fig_5}.}
			\label{fig_6}
		\end{center}
	\end{figure}

	Although STM measurements in this type of samples would remain a practical challenge due to the encapsulation with the top hBN layer, it is 
	interesting to discuss the SM effects on the spatial distribution of the LDOS (SLDOS). As expected, SLDOS is modulated 
	by the hexagonally symmetric potential which originates from moir\'e patterns with different spatial lengths. In Fig.~\ref{fig_6}(a) 
	clear modulation respecting the symmetry of the super-moir\'e can be observed. The energy is just above the SM induced Dirac-like point, where the modulation is 
	the strongest. Notice that the highest LDOS appears in the corners of the SM unit-cell, where graphene atoms are maximally covered with boron and nitrogen atoms 
	from both the top and the bottom hBN. Differently, in Figs.~\ref{fig_6}(b-d) close to the $\theta_1$ and $\theta_2$ moir\'e SDPs, different symmetry of 
	the SLDOS patterns reveals the dominant effect of the two underlying moir\'es. In this case, the LDOS distribution at the corners of the SM is minimized. 

\section{Enhancement of super-moir\'e by lattice relaxation}
\label{Sec:relaxation}

	When graphene and hBN are put on top of each other and aligned, two effects occur. Firstly there is band energy reconstruction through the 
	appearance of mini-bands and of secondary Dirac points. The second effect is related to the van der Waals energy which is minimized through the 
	formation of stacking areas with minimal energy~\cite{woods2014commensurate}. This induces strain which further changes the tunnelling 
	probability (or the hopping energy) throughout the sample. In the case of incommensurate alignment of layers, when no or little relaxation occurs, 
	the effects of the opposite boron and nitrogen species on the graphene layer are averaged out on the area of the moir\'e unit cell, 
	and no gaps	appear~\cite{woods2014commensurate}. 

	Relaxation changes the interatomic registry, which results in the appearance of significant gaps at the primary and secondary Dirac 
	points~\cite{jung2015origin, kim2018accurate}, which was confirmed experimentally
	~\cite{woods2014commensurate, hunt2013massive, chen2014observation, finney_tunable_2019, yang2013epitaxial, yankowitz2018dynamic}. 

	If the goal is to preserve the gapless linear spectrum, an addition of misaligned top layer to the hBN substrate, 
	would again average out the mass-like boron/nitrogen terms and preserve the gapless graphene spectrum. More interesting effects come into play
	when the second hBN layer is close to align~\cite{finney_tunable_2019}.
		
\begin{figure}[!htb]
		\begin{center}
			\includegraphics[width=\linewidth]{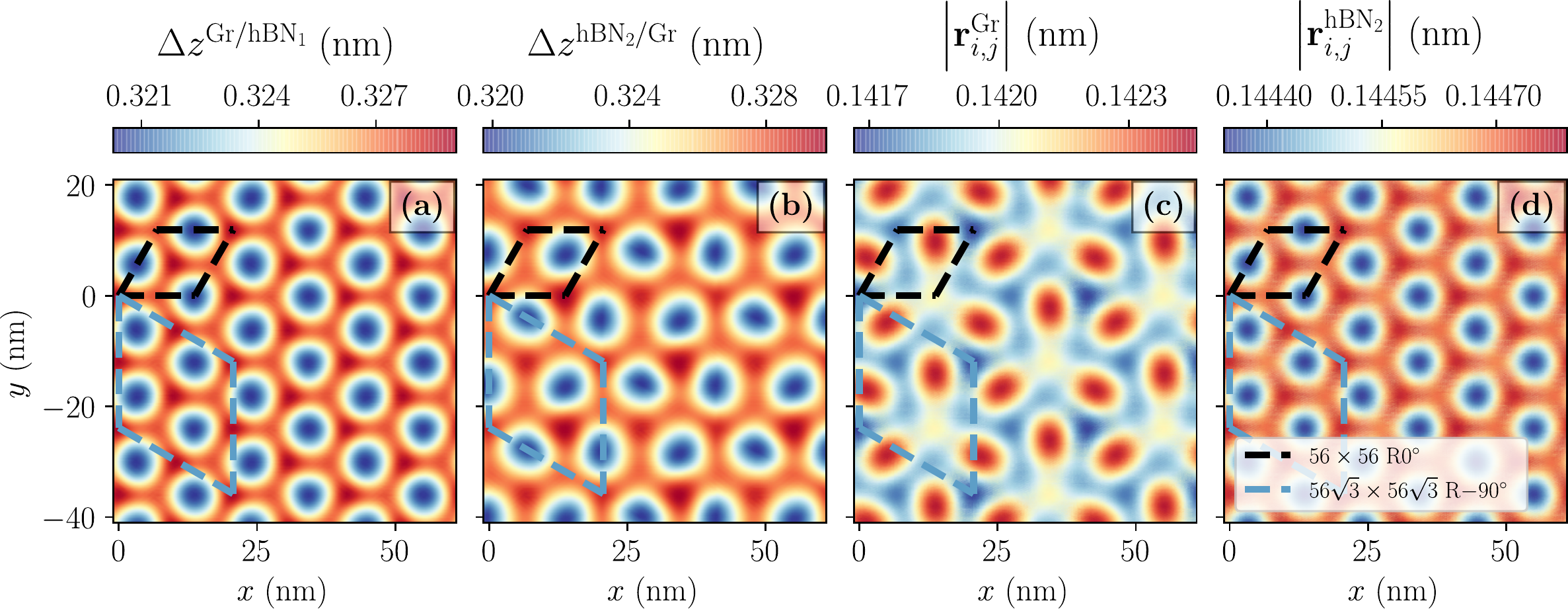}
			\caption{Relaxation of trilayer hBN/gr/hBN with $\theta_1= 0^\circ$ $\theta_2\approx 0.6^\circ$. Interlayer distance between graphene and (a) bottom and (b) top hBN layer.
			In panels (c) and (d) bond lengths in relaxed graphene and top hBN layer are shown, respectively.}
			\label{fig_7}
		\end{center}
	\end{figure}

	The sample relaxation was done using MD simulations. Bottom hBN layer is kept rigid to mimic the effect of a thick substrate, while graphene and top hBN 
	are freely relaxed. The resulting effect on the interlayer distance ($\Delta z$) and the bond lengths ($\mathbf{r}_{i,j}$) is shown in 
	Fig.~\ref{fig_7}. Interlayer distance between graphene and bottom hBN in Fig.~\ref{fig_7}(a) shows a clear evidence of the symmetry which follows 
	the moir\'e pattern between these two layers, as marked with the black unit cell. Notice also that the bond lengths in 
	top	hBN reveal the symmetry of the moir\'e between graphene and this layer shown in (d). In contrast, the distance between top hBN and graphene in panel (b) and 
	the distribution of the bond lengths in graphene in panel (c) follow the symmetry of the SM, marked with the blue unit cell. This shows that the SM is 
	imprinted in the in-plane distribution of the strain in graphene, and the interlayer distance between graphene and the relaxed hBN layer, modifying in such a 
	way all the hopping terms in Eq.~(\ref{equation:hoppings}) and further enhancing the super-moir\'e effect. 

\begin{figure}[!htb]
		\begin{center}
			\includegraphics[width=\linewidth]{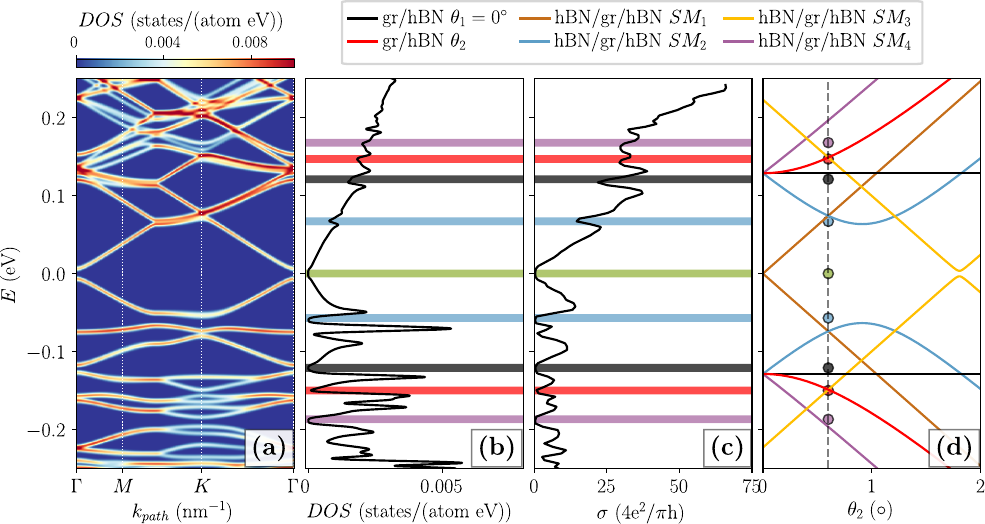}
			\caption{Electronic properties of relaxed hBN/gr/hBN system with $\theta_1= 0^\circ$ and $\theta_2\approx0.6^\circ$, corresponding to the unrelaxed system of
			Fig.~\ref{fig_5}. (a) Band structure 
			along the $\Gamma - M - K - \Gamma$ path of the super-moiré mini-Brillouin zone. (b) DOS, and (c) conductivity. 
			Dominant super-moir\'{e} feature is marked with blue colour, black marks the aligned $\theta_1$ hBN/gr, and red 
			rotated $0.6^\circ$ hBN/gr features. Purple marks the higher order super-moir\'{e} feature. Green line marks the primary 
			graphene Dirac point. (d) Expected positions of the super-moir\'{e} features using $v_F\approx 0.75\times10^6$ m/s.}
			\label{fig_8}
		\end{center}
	\end{figure}

	In the electronic spectrum, both the rigid, Fig.~\ref{fig_5}(a), and the relaxed sample, Fig.~\ref{fig_8}(a), 
	show similar behaviour around the two SDPs at the hole side. The main difference is that in the relaxed sample gaps appear 
	at the primary (see figure S5 in~\cite{Zihao2019}), and in both of the secondary Dirac points on the hole doped side~\cite{kim2018accurate}, consistent with broken 
	inversion symmetry where the gap does not average out over the area of the SM pattern. It is important to note that relaxation effects are believed to be only 
	partially responsible for the appearance of the experimentally observed gaps, while they are expected to increase when many-body interactions are included
	~\cite{hunt2013massive, jung2015origin, jung2017moire_gaps}.

	More importantly, the band structure in Fig.~\ref{fig_8}(a) reveals a gap at the hole doped SM Dirac-like point as well. The appearance of the SM 
	reconstruction is highly tunable within the low energy spectrum through the change of the twist angle. In the small misalignment regime,
	the minimization of the van der Waals energy induces surface reconstruction that further results in SM gaps. Hence, the hBN/gr/hBN trilayers might provide a 
	feasible route for gap engineering at arbitrary energies, which was not possible previously in moir\'e like structures. Simply, the lowest energy 
	SDP ($E_{SDP}\approx \pm140$ meV) is limited by the largest bilayer hBN/gr period which is $\lambda\approx 13.9$ nm. 

	The slight misalignment brings the two SDPs to a close proximity, which then results in the appearance of a narrow band in between the two
	energy features, $E_{SDP1}$ and $E_{SDP2}$~\cite{finney_tunable_2019}. Due to relaxation of the sample, this band becomes even narrower. 
	In addition, the SM gap edge shows similar dispersionless property, which was absent in the rigid sample. This means that by relaxing the 
	sample, two almost flat	bands appear, which suggests even richer physics in this regime.

	Relaxation partially recovers the electron-hole symmetry in the spectrum. Although gaps appear only in the hole doped side, the moir\'e and SM become 
	more visible on the electron side, as shown in the DOS, Fig.~\ref{fig_8}(b), and the conductivity, Fig.~\ref{fig_8}(c).  

\section{Conclusion}
\label{sec:conclusion}
	To conclude, we have examined and showed the effects of slightly misaligned hBN encapsulation layers on the electronic properties of monolayer graphene.
	A simple model in terms of Bragg scattering vectors is able to explain the origin of the super-moir\'e spectrum. 
	Using a real space tight-binding method we have quantitatively described the angle dependent SM effects that result in additional spectrum reconstruction,
	velocity suppression and spatial electron redistribution. 

	In addition to rigid alignment between the layers, we have studied the effects of lattice relaxation. Bottom hBN layer was kept rigid, which simulates 
	the effect of a thick substrate, while the top hBN and graphene were freely relaxed. By relaxing the trilayer sample, the SM features are enhanced,
	and multiple asymmetrical low-energy gaps and flat bands whose energy can be controlled by misalignment appear.	The opening of band-gaps, and the expected 
	enhancement due to many-body interactions appearing alongside flat bands are of high interest to the community working on 
	twisted 2D (hetero)structures. The fact that flat bands are apparent in the case of graphene almost aligned to two encapsulated hBN layers, means only that 
	further theoretical studies starting from the picture of interacting electrons are needed to understand all the different aspects of close aligned hBN/gr/hBN layers. 

	Finally, this study agrees well with recent experimental studies~\cite{Lujun_2019, finney_tunable_2019, Zihao2019} adding to the understanding of more subtle effects of complex moir\'e patterns that can appear when 
	combining hetero-multi-layers.

\section*{Acknowledgements}
	This work was funded by FLAGERA project TRANS2DTMD and the Flemish Science Foundation (FWO-Vl) through a postdoc fellowship for S.P.M. The authors acknowledge useful discussions with 
	W. Zihao and K. Novoselov. 

\section*{References}
\bibliographystyle{iopart-num}
\bibliography{bibliography}

\pagebreak
\begin{center}
\textbf{\large Supplementary information: Double moir\'{e} with a twist : super-moir\'e in encapsulated graphene.}
\end{center}
\setcounter{equation}{0}
\setcounter{figure}{0}
\setcounter{table}{0}
\setcounter{section}{0}
\setcounter{page}{1}
\makeatletter
\renewcommand{\theequation}{S\arabic{equation}}
\renewcommand{\thefigure}{S\arabic{figure}}
\renewcommand{\thesection}{S\arabic{section}}

\author{M. An\dj elkovi\'{c}, S. P. Milovanovi\'{c}, L. Covaci, and F. M. Peeters}

\address{Departement Fysica, Universiteit Antwerpen, Groenenborgerlaan 171, B-2020 Antwerpen, Belgium}
\ead{misa.andelkovic@uantwerpen.be, slavisa.milovanovic@uantwerpen.be, lucian.covaci@uantwerpen.be, francois.peeters@uantwerpen.be}
\vspace{10pt}
\begin{indented}
	\item[]\today
\end{indented}

\section{Identifying the rotation angle}
\label{sec:Identifying_rotation}

\begin{figure}[!htb]
		\begin{center}
			\includegraphics[width=\linewidth]{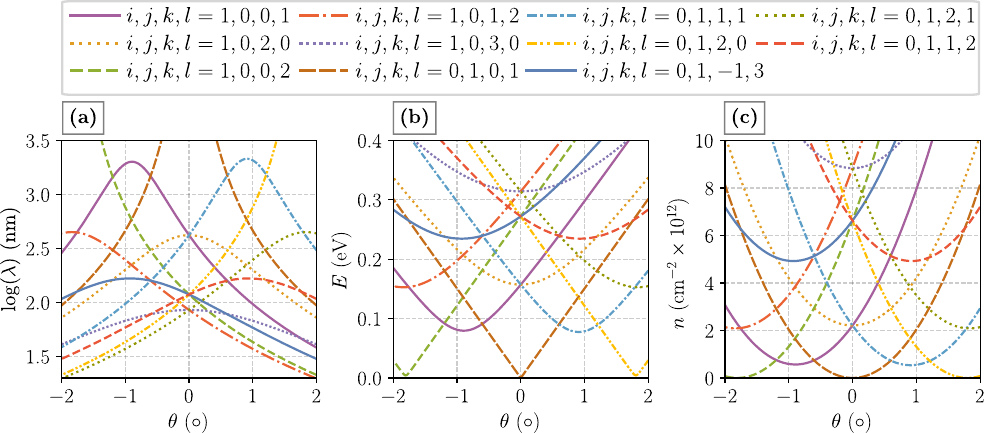}
			\caption{Super-moir\'e (a) period, (b) energy and (c) the density corresponding to different harmonics as defined in Eq.~(\ref{eq:define_arb_super_moire}).}
			\label{fig_s1}
		\end{center}
	\end{figure}

	Using geometrical considerations explained in the main text we are able to relate the theoretical prediction of the super-moir\'e harmonics to 
	real-space periods, reconstruction energies or densities. Different combinations of moir\'e vectors give more harmonics in addition to the harmonics 
	already defined in Eq.~(\ref{eq:sm_harmonics}). In Fig.~\ref{fig_s1}, panels (a-c) period, reconstruction energy, and density are shown for $11$ different 
	harmonics that appear in the low energy spectrum (lower than $0.4$ eV). Experimental results are obtained mostly as a function of carrier density,
	so panel~\ref{fig_s1}(c) can be particularly important for extracting the expected values of the rotation angles, and the period of the 
	underlying periodicities from the super-moir\'e features.

\section{Band structure - comparison with aligned bilayer}
\label{sec:bs_comparison}

\begin{figure}[!htb]
		\begin{center}
			\includegraphics[width=\linewidth]{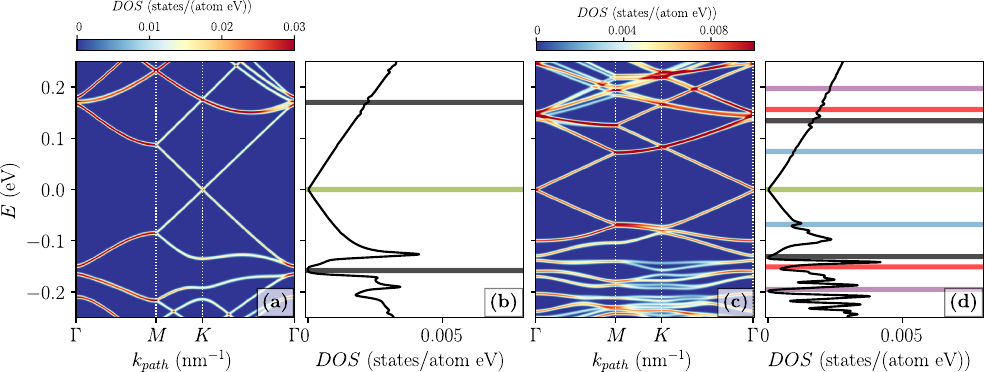}
			\caption{Band structure and DOS of (a, b) bilayer hBN/gr $\theta_1=0^\circ$ and (c, d) trilayer hBN/gr/hBN for $\theta_1=0^\circ$ and 
			$\theta_2\approx0.6^\circ$.}
			\label{fig_s2}
		\end{center}
	\end{figure}

	By solving the Diophantine equation from Eq.~(\ref{eq:diophantine}), commensurate structures that allow to define a unit cell can be obtained.
	One of the solutions is for rotations $\theta_1=0^\circ$ (aligned hBN) and $\theta_2\approx 0.6^\circ$ (rotated hBN) resulting in 
	the super-moir\'{e} period of $\approx 23.8$ nm. Figure~\ref{fig_s2} shows the band structure along the path $\Gamma - M - K - \Gamma$ 
	in the SM mini-Brillouin zone and the DOS for a bilayer (aligned hBN/gr $\theta_1=0^\circ$) and trilayer hBN/gr/hBN
	($\theta_1=0^\circ$ and $\theta_2\approx0.6^\circ$), respectively.

	In the bilayer case (Figs.~\ref{fig_s2}(a, b)), the hBN substrate induces a pair of secondary Dirac points at energies 
	$\approx \pm 140$ meV, with large degree of electron-hole asymmetry which was discussed previously (at energies marked by black line), 
	where $E_{DP}$ is the (primary) Dirac point of graphene (marked by green line).
	The additional top hBN layer (Figs.~\ref{fig_s2}(c, d)), as suggested from the other spectral calculations, results in additional Dirac-like points 
	(marked by blue line in the DOS) at energies below the ones induced by aligned or rotated 2 layer moir\'{e} patterns 
	($\sim \pm 70$ meV). Additional secondary Dirac point due to the rotated hBN/gr appears as well 
	(marked by red line).

	The Dirac point in graphene is located at different points in the moir\'e or super-moir\'e mini-Brillouin zone, which is caused by different superlattice 
	band folding~\cite{Zhou2013}. 

\section{Behaviour at higher rotation angles}
\label{sec:higher_rotation_angles}

\begin{figure}[!htb]
		\begin{center}
			\includegraphics[width=\linewidth]{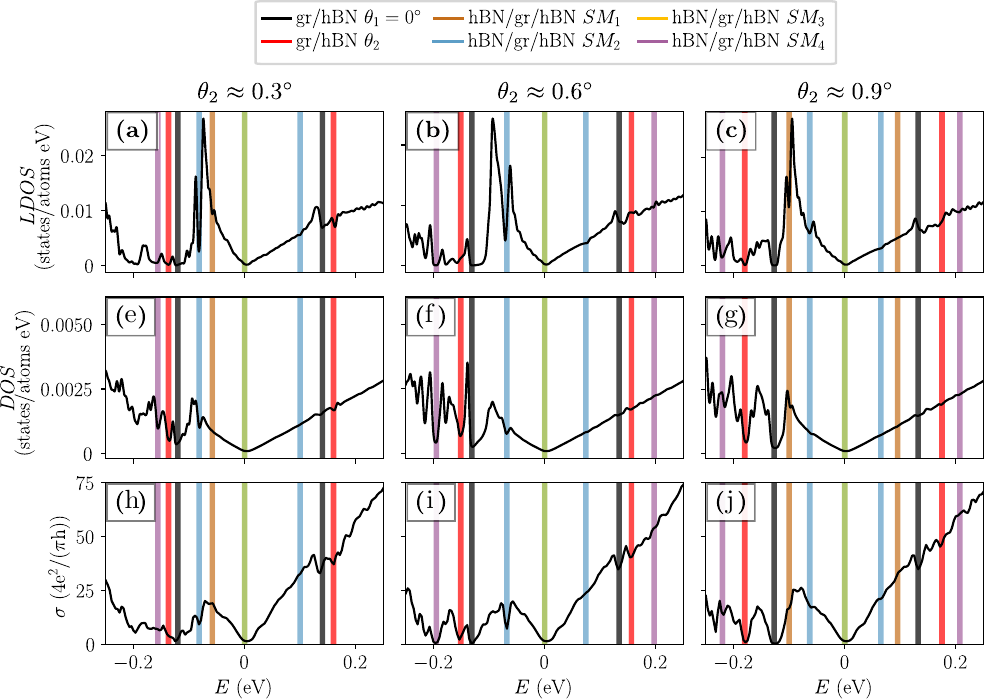}
			\caption{Cuts of LDOS from Fig.~\ref{fig_4}(b) for $\theta_2\approx 0.3$, $\theta_2\approx 0.6$, $\theta_2\approx 0.9$, panels (a-c),
			together with the conductivity (e-g) and the DOS plots in panels (h-j). Different coloured stripes in each figure correspond to fits of the harmonics defined in 
			the legend.}
			\label{fig_s3}
		\end{center}
	\end{figure}

	To expand the discussion on the SM reconstruction effects, in Fig.~\ref{fig_4} LDOS (at the same graphene site as in Fig.~\ref{fig_4}(b), panels~\ref{fig_s3}(a-c)), 
	DOS (panels~\ref{fig_s3}(e-g)) and conductivity (panels~\ref{fig_s3}(h-j)) are shown in a system with rotation angles $\theta_1=0^\circ$ and $\theta_2=0.3^\circ$, 
	$\theta_2=0.6^\circ$, $\theta_2=0.9^\circ$. The fits of the SM effects are marked with coloured stripes, where the different harmonics are labeled in the legend. 
	Notice that the effects of the SM are more apparent in the LDOS when compared to global properties like DOS and conductivity, where the effects are averaged over 
	different stacking regions. The SM features are almost completely wiped out at rotation $\theta_2\approx 0.9^\circ$, which confirms our findings about the necessity 
	of close alignment of the layers in order to see the SM effects.

\end{document}